\documentclass[twocolumn,showpacs,jcp,amsfonts,amsmath,amssymb,floatfix]{revtex4} 
\usepackage{color}
\usepackage{hhline}
\usepackage{mathrsfs}
\usepackage{graphicx}
\usepackage{dcolumn}
\usepackage{bm}
\usepackage{multirow}
\usepackage{booktabs}
\usepackage{afterpage}
\usepackage{amsmath}

\begin{document}

\title{Iterative diagonalization of the non-Hermitian transcorrelated Hamiltonian using a plane-wave basis set: Application to $sp$-electron systems with deep core states}
\author{Masayuki Ochi$^{1,2}$}
\email{ochi@phys.sci.osaka-u.ac.jp}
\altaffiliation[Present address: ]{Department of Physics, Osaka University, Machikaneyama-cho, Toyonaka, Osaka 560-0043, Japan}
\author{Yoshiyuki Yamamoto$^{3}$}
\author{Ryotaro Arita$^{1,2}$}
\author{Shinji Tsuneyuki$^{3,4}$}
\affiliation{$^1$RIKEN Center for Emergent Matter Science (CEMS), Wako, Saitama 351-0198, Japan}
\affiliation{$^2$JST ERATO Isobe Degenerate $\pi$-Integration Project, Advanced Institute for Materials Research (AIMR), Tohoku University, Katahira, Aoba-ku, Sendai, Miyagi 980-8577, Japan}
\affiliation{$^3$Department of Physics, The University of Tokyo, Hongo, Bunkyo, Tokyo 113-0033, Japan}
\affiliation{$^4$Institute for Solid State Physics, The University of Tokyo, Kashiwa, Chiba 277-8581, Japan}

\date{\today}
\begin{abstract}
We develop an iterative diagonalization scheme in solving a one-body self-consistent-field equation in the transcorrelated (TC) method using a plane-wave basis set.
Non-Hermiticity in the TC method is well handled with a block-Davidson algorithm. We verify the required computational cost is efficiently reduced by our algorithm.
In addition, we apply our plane-wave-basis TC calculation to some simple $sp$-electron systems with deep core states to elucidate an impact of the pseudopotential approximation to the calculated band structures. We find a position of the deep valence bands is improved by an explicit inclusion of core states, but an overall band structure is consistent with a regular setup that includes core states into the pseudopotentials. This study offers an important understanding for the future application of the TC method to strongly correlated solids.
\end{abstract}

\maketitle

\section{Introduction}

Density functional theory (DFT)~\cite{HK, KS} has provided valuable information of the electronic structure for a wide range of materials over several decades.
However, it is now well known that popular approximations in DFT such as local density approximation (LDA) and generalized gradient approximation (GGA) lack accuracy for strongly correlated systems.
One possible strategy to go beyond is to develop highly accurate DFT functionals.
Such trials have improved the situation to some extent, but the exact form of the exchange-correlation energy functional is out of reach owing to its non-trivial nature.

Another strategy is to employ a promising theoretical framework independent of DFT: wave function theory (WFT) such as quantum chemical methods~\cite{Szabo, FCIQMC} and several flavors of novel quantum Monte Carlo (QMC) methods~\cite{QMCreview, FCIQMC}.
WFT explicitly handles a many-body wave function, which enables systematic improvement of accuracy.
Although this strategy works well for molecular systems, we have been faced with expensive computational cost for solids as a result of such an explicit treatment of many-body problems.
From this viewpoint, the transcorrelated (TC) method~\cite{BoysHandy, Handy, Ten-no1, Umezawa}, one of WFTs, has been attracting much attention because of its fascinating advantages with practical computational cost.
The TC method makes use of a single-particle picture under an effective Hamiltonian explicitly represented with the Jastrow correlation factor.
For molecular systems, some recent works have shown a high potentiality of the TC method~\cite{Ten-no1,Ten-no2,Ten-no3}.
Optimization of the wave function with a help of Monte Carlo integration~\cite{Umezawa,LuoTC,LuoVTC} and development of the canonical TC theory~\cite{CanonicalTC}
are also remarkable recent innovation.
For solid-states calculations, the TC method is shown to reproduce accurate band structures~\cite{Sakuma, TCaccel, TCjfo} and optical absorption spectra~\cite{TCCIS}.
The TC method and a related theory have been successfully applied also to the Hubbard model~\cite{TCHubbard,LieAlgebra}.
It is noteworthy that the TC method has a close relationship with some QMC methods in the sense that both methods employ the Jastrow correlation factor.

As of now, the TC method has been applied only to weakly correlated systems in solid-states calculations. 
To apply the TC method to strongly correlated solids, it is necessary to employ an efficient basis set for representing the one-electron orbitals in the TC method.
In the former studies~\cite{Sakuma, TCaccel}, the TC orbitals are expanded with the LDA orbitals.
Whereas a small number of LDA orbitals is required for convergence in weakly correlated systems, its number will inevitably increase when one handles localized states, which are not well-described by LDA. As a matter of fact, we shall see that an enormously large number of LDA orbitals are sometimes required and then computational requirement becomes very expensive. This problem should be resolved for further advance in the TC method, especially for application to strongly correlated systems. Note that because an effective Hamiltonian in the TC method is non-Hermitian, whether an efficient convergence is achieved in the TC method just like other conventional methods is a non-trivial problem. 

In this study, we develop an iterative diagonalization scheme in solving a one-body self-consistent-field (SCF) equation in the TC method using a plane-wave basis set.
We find that a subspace dimension to represent the TC orbitals drastically decreases compared with the former LDA-basis approach. 
This new development enhances applicability of the TC method to various systems.
As a test, we apply our new algorithm to some simple $sp$-electron systems with deep core states (i.e., a part of the deep core states are not just included in pseudopotentials but explicitly treated in a many-body wave function) and clarify how the core electrons affect the TC band structures.
Treatment of core electrons has recently been recognized as an important issue to be taken care of.
In the {\it GW} method~\cite{GW1,GW2,GW3}, it has been pointed out that an error arising from the pseudopotential is not negligible in some cases~\cite{GWpp1,GWpp2,GWpp3,GWpp4,GWpp5,GWpp6,GWpp7,GWpp8}.
The same situation was reported also in QMC calculations~\cite{QMCreview,QMCpp1,QMCpp2}.
In the TC method, such an effect has not been investigated, but to elucidate the impact of core electrons on the band structure is an important and unavoidable problem.
In addition, because the TC method employs a many-body wave function common to some QMC methods, such investigation will provide a valuable information also for the QMC community.
We find that an explicit treatment of core states improves a position of deep valence states whereas the band structures in upper energy region do not exhibit large changes, which means that our choice of the Jastrow factor can provide consistent electronic structures for a wide energy range whether the core electrons are explicitly treated or not.
Our findings in this study encourage application of the TC method to strongly correlated materials with $d$ electrons where an explicit treatment of semi-core $sp$ electrons with the same principal quantum number will be necessary~\cite{GWpp8,QMCpp1}.

The present paper is organized as follows. In Sec.~\ref{sec:2}, we give a brief introduction of basic features of the TC method.
In Sec.~\ref{sec:3}, we present a block-Davidson algorithm to solve a one-body SCF equation in the TC method using a plane-wave basis set.
In Sec.~\ref{sec:4}, we demonstrate how efficiently our new algorithm works, and using this algorithm, we compare the band structures calculated with core electrons explicitly treated in many-body wave functions and those just included in pseudopotentials for some simple $sp$-electron systems.
Section~\ref{sec:5} summarizes this study.

\section{Transcorrelated method}\label{sec:2}

Since a detailed theoretical framework of the TC method was presented in previous papers~\cite{BoysHandy, Ten-no1, Umezawa}, we just make a brief review here.
In electronic structure calculations, our objective is to solve the Schr{\" o}dinger equation
\begin{equation}
\mathcal{H}\Psi = E\Psi,
\end{equation}
for the many-body Hamiltonian $\mathcal{H}$ under the external potential $v_{\mathrm{ext}}(x)$,
\begin{equation}
\mathcal{H}=\sum_{i=1}^{N} \left( -\frac{1}{2}\nabla_i ^2 + v_{\mathrm{ext}}(x_i) \right) +
\frac{1}{2} \sum_{i=1}^{N}\sum_{j=1(\neq i)}^N \frac{1}{|\mathbf{r}_i-\mathbf{r}_j|},\label{eq:Hamil}
\end{equation}
where $x=(\mathbf{r},\sigma)$ denotes a set of spatial and spin coordinates associated with an electron. 
First, we formally factorize the many-body wave function $\Psi$ as $\Psi=F\Phi$ where $F$ is the Jastrow factor
\begin{equation}
F=\mathrm{exp}(-\sum_{i,j(\neq i)=1}^N u(x_i,x_j)),
\end{equation}
and $\Phi$ is defined as $\Psi/F$. Next, we perform a similarity transformation of the Hamiltonian as
\begin{equation}
\mathcal{H}\Psi = E\Psi \Leftrightarrow \mathcal{H}_{\mathrm{TC}}\Phi = E \Phi \ \ \ (\mathcal{H}_{\mathrm{TC}} =F^{-1}\mathcal{H}F).
\end{equation}
In this way, the electron correlation described with the Jastrow factor is incorporated into
the similarity-transformed Hamiltonian $\mathcal{H}_{\mathrm{TC}}$, which we called the TC Hamiltonian hereafter.
A Jastrow function $u$ is chosen as the following simple form:~\cite{TCjfo,QMCreview}
\begin{equation}
u(x,x')=\frac{A}{|\mathbf{r}-\mathbf{r'}|}
\left( 1-\mathrm{exp}\left( -\frac{|\mathbf{r}-\mathbf{r'}|}{C_{\sigma,\sigma'} } \right) \right) ,
\label{eq:Jastrow}
\end{equation}
where
\begin{align}
A&=\sqrt{\frac{V}{4\pi N}}\times \sqrt{1-\frac{1}{\varepsilon}}, \label{eq:JastrowA} \\
C_{\sigma, \sigma'} &= \sqrt{2A}\ (\sigma=\sigma'), \sqrt{A}\ (\sigma\neq\sigma'),
\end{align}
with $N$, $V$, and $\varepsilon$ being the number of electrons in the simulation cell, the volume of the simulation cell, and the static dielectric constsant, respectively.
Core electrons effectively included in pseudopotentials are not counted in the definition of $N$, which should be the same as that in Eq.~(\ref{eq:Hamil}).
A choice of $\varepsilon$ in this study shall be described in Sec.~\ref{sec:4A}.
The asymptotic behavior of this function is determined so as to reproduce the screened electron-electron interaction in solids $1/(\varepsilon r)$ in the long-range limit~\cite{BohmPines, TCjfo} and satisfy the cusp condition~\cite{cusp,cusp2} in the short-range limit.
To satisfy the exact cusp condition for singlet and triplet pairs of electrons, one should adopt the operator representation of the Jastrow function, which introduces troublesome non-terminating series of the effective interaction in the TC Hamiltonian~\cite{Ten-nocusp}. Thus we use the approximated cusp conditions in the same manner as in QMC studies~\cite{QMCreview}.
Because this Jastrow function captures both the short- and long-range correlations, a mean-field approximation for the TC Hamiltonian is expected to work well.
Here we approximate $\Phi$ to be a single Slater determinant consisting of one-electron orbitals: $\Phi=\mathrm{det}[ \phi_i(x_j) ]$, and then a one-body SCF equation is derived:
\begin{align}
\left( -\frac{1}{2}\nabla_1^2 +v_{\mathrm{ext}}(x_1) \right) \phi_i (x_1)\notag \\
+ \sum_{j=1}^N
\int \mathrm{d}x_2\  \phi_j^*(x_2) v_{\mathrm{2body}}(x_1,x_2)
\mathrm{det} \left[
\begin{array}{rrr}
\phi_i(x_1) & \phi_i(x_2) \\
\phi_j(x_1) & \phi_j(x_2) \\
\end{array} \right] \notag \\
- \frac{1}{2}\sum_{j=1}^N \sum_{k=1}^N
\int \mathrm{d}x_2 \mathrm{d}x_3\  \phi_j^*(x_2)\phi_k^*(x_3)v_{\mathrm{3body}}(x_1,x_2,x_3)  \notag \\
\times 
\mathrm{det} \left[
\begin{array}{rrr}
\phi_i(x_1) & \phi_i(x_2) &  \phi_i(x_3) \\
\phi_j(x_1) & \phi_j(x_2) & \phi_j(x_3) \\
\phi_k(x_1) & \phi_k(x_2) & \phi_k(x_3)
\end{array} \right]
= \sum_{j=1}^N \epsilon_{ij} \phi_j(x_1), \label{eq:SCF}
\end{align}
where $v_{\mathrm{2body}}(x_1,x_2)$ and $v_{\mathrm{3body}}(x_1,x_2,x_3)$ are the effective interactions in the TC Hamiltonian defined as
\begin{align}
v_{\mathrm{2body}}(x_1,x_2)\notag\\
\equiv \frac{1}{|\mathbf{r}_1-\mathbf{r}_2|}+\frac{1}{2}\big(\nabla_1^2 u(x_1,x_2)+\nabla_2^2 u(x_1,x_2)\notag \\
-(\nabla_1 u(x_1,x_2))^2-(\nabla_2 u(x_1,x_2))^2\big) \notag \\
+ \nabla_1 u(x_1,x_2)\cdot \nabla_1 + \nabla_2 u(x_1,x_2)\cdot \nabla_2,
\end{align}
and
\begin{align}
v_{\mathrm{3body}}(x_1,x_2,x_3)\notag\\
\equiv\nabla_1 u(x_1,x_2)\cdot \nabla_1 u(x_1,x_3) 
+ \nabla_2 u(x_2,x_1) \cdot \nabla_2 u(x_2,x_3) \notag \\
+ \nabla_3 u(x_3,x_1) \cdot \nabla_3 u(x_3,x_2).
\end{align}
By solving Eq.~(\ref{eq:SCF}), one can optimize TC one-electron orbitals.
This procedure costs just the same order as the uncorrelated Hartree-Fock (HF) method with a help of an efficient algorithm~\cite{TCaccel}.
By construction, $\Phi$ can be systematically improved over a single Slater determinant~\cite{Ten-no1,Ten-no2,Ten-no3,TCCIS,TCMP2}, which is an important advantage of the TC method, but in this study, we focus on the case where $\Phi$ is supposed to be a single Slater determinant.

We note that the non-Hermiticity of the TC Hamiltonian originating from the non-unitarity of the Jastrow factor is essential in the description of the electron correlation effects. It is obvious that the purely imaginary Jastrow function $u(x_i, x_j)$ that makes the Jastrow factor exp[$-\sum_{i,j(\neq i)} u(x_i, x_j)$] unitary cannot fulfill the cusp condition and the long-range asymptotic behavior mentioned before.  Although some previous studies adopted approximations for the effective interaction in the TC Hamiltonian to restore the Hermiticity for molecular systems with a small number of electrons~\cite{LuoVTC,CanonicalTC}, it is unclear that such approximation is valid for general molecular and periodic systems. Therefore, in this study, we explicitly handle the non-Hermiticity of the TC Hamiltonian without introducing additional approximations.

We also mention the bi-orthogonal formulation of the TC method, which we called the BiTC method.
The BiTC method was applied to molecules~\cite{Ten-no2} and recently also to solids~\cite{TCMP2}.
A detailed description of the BiTC method can be found in these literatures.
In the BiTC method, we use left and right Slater determinants consisting of different one-electron orbitals: $X=\mathrm{det}[\chi_i(x_j)]$ and $\Phi=\mathrm{det}[\phi_i(x_j)]$, respectively, with the bi-orthogonal condition $\langle \chi_i | \phi_j \rangle = \delta_{i,j}$ and the normalization condition $\langle \phi_i | \phi_i \rangle = 1$. Then a one-body SCF equation becomes slightly different from Eq.~(\ref{eq:SCF}) in the sense that `bra' orbitals $\phi^*(x)$ are replaced with $\chi^*(x)$.
Because the similarity transformation of Hamiltonian introduces non-Hermiticity, such formulation yields a different result from the ordinary TC method.
In this study, we investigate both the TC and BiTC methods.

\section{Block-Davidson algorithm for plane-wave calculation}\label{sec:3}

In the former studies of the TC method applied to solid-state calculations, we have used LDA orbitals as basis functions to expand the TC orbitals in solving the SCF equation for efficient reduction of computational cost~\cite{Sakuma}.
This prescription works well if a moderate number of LDA orbitals are required for convergence as in previous studies.
However, this is not necessarily the case when we deal with the systems where TC and LDA orbitals are expected to exhibit sizable differences, e.g., for strongly correlated systems.
To overcome this problem, we develop an iterative diagonalization scheme using a plane-wave basis set.
Because the TC Hamiltonian is non-Hermitian owing to the similarity transformation, some standard methods such as the conjugate gradient (CG) method~\cite{CG} do not work and so we adopted the block-Davidson algorithm~\cite{Davidson1,Davidson2}.
Whereas the block-Davidson algorithm has been successfully applied to other conventional methods such as DFT, some modifications described below are necessary to adopt it to the TC method.

\begin{figure}
 \begin{center}
  \includegraphics[scale=.3]{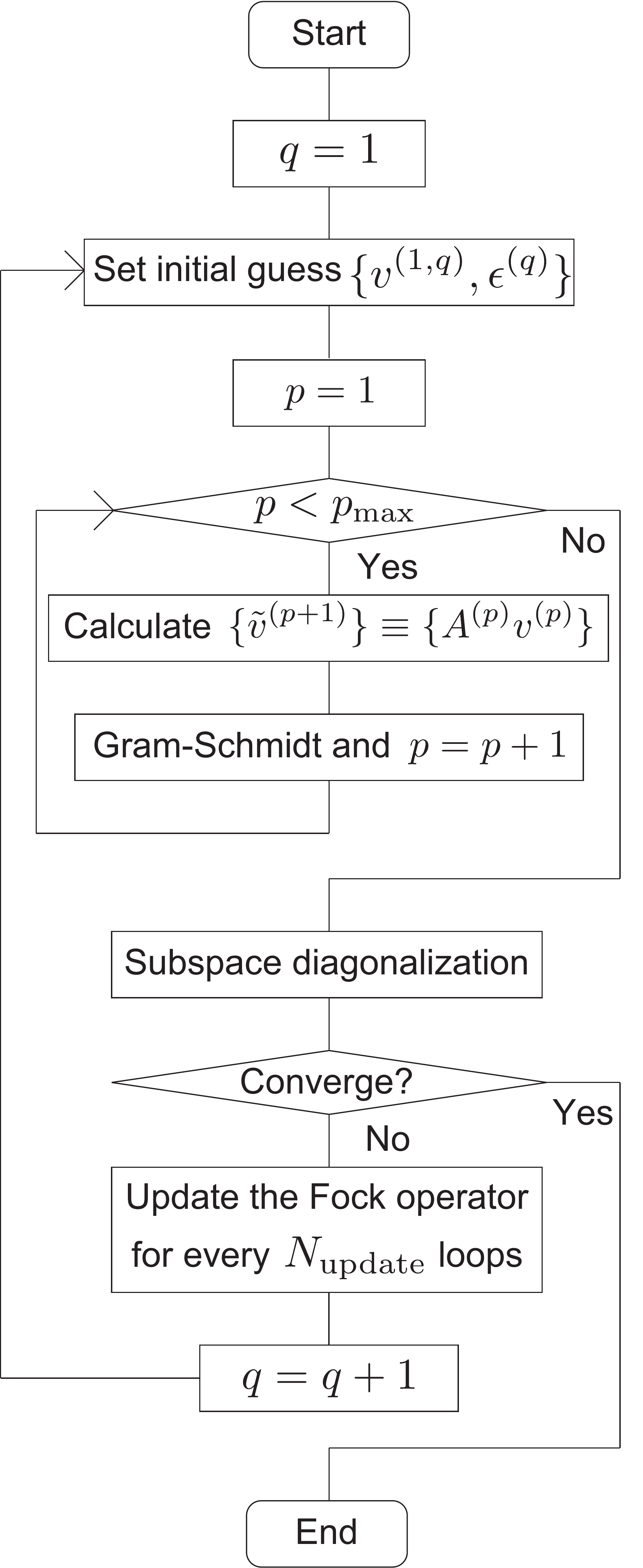}
  \caption{Flow of the iterative diagonalization scheme for solving the SCF equation in the TC method using a plane-wave basis set.}
  \label{fig:Algo}
 \end{center}
\end{figure}

Figure~\ref{fig:Algo} presents a flow of our calculation.
Here we define the TC-Fock operator $F[\phi]$ as $F[\phi]\phi_i(x)$ equals the left-hand side of Eq.~(\ref{eq:SCF}).
Our algorithm consists of a double loop.
In the inner loop labeled with $p$, a subspace dimension is gradually increased.
In the outer loop labeled with $q$, the TC-Fock operator is diagonalized in that subspace and the convergence is checked. Both indices start from $1$. Detailed description of the algorithm is presented below.

\subsection{Inner loop: subspace extension}

First, we begin with the initial trial vectors $\{ v_1^{(1,q)}, v_2^{(1,q)}, \dots, v_n^{(1,q)} \}$ and eigenvalues $\{ \epsilon_1^{(q)},\epsilon_2^{(q)},\dots,\epsilon_n^{(q)} \}$ of the TC-Fock operator, i.e. the initial estimates of the TC orbitals $\phi$ and the eigenvalues of the $\epsilon_{ij}$ matrix in Eq.~(\ref{eq:SCF}), at each $k$-point $\mathbf{k}$ where $n$ is the number of bands to be calculated here. These initial estimates are obtained in the previous $(q-1)$-th outer loop, where $\langle v_i^{(1,q)} | v_j^{(1,q)} \rangle = \delta_{ij}$ ($1\leq i, j \leq n$) is satisfied.
In the first loop, i.e. $q=1$, we set LDA orbitals and their energies as initial guesses.
Next, we calculate $\{ \tilde{v}_1^{(p+1,q)}, \tilde{v}_2^{(p+1,q)},\dots, \tilde{v}_n^{(p+1,q)} \} \equiv \{ A_1^{(p,q)}v_1^{(p,q)}, A_2^{(p,q)}v_2^{(p,q)},\dots, A_n^{(p,q)}v_n^{(p,q)} \}$ where
\begin{equation}
A_m^{(p,q)}v_m^{(p,q)}(\mathbf{G})\equiv T_m^{(p,q)}(\mathbf{G})(F[v^{(1,\tilde{q})}]-\epsilon_m^{(q)})v_m^{(p,q)}(\mathbf{G})
\end{equation}
for the wave vector $\mathbf{G}$ and the preconditioner $T_m^{(p,q)}(\mathbf{G})$. $\tilde{q}$ is defined later.
The preconditioner is introduced to reduce numerical errors by filtering high-energy plane waves.
Diagonal elements of the Fock operator in terms of a plane-wave basis set are usually used in the preconditioner, but their evaluation requires sizable computational effort for the TC method owing to the existence of the three-body terms.
Thus we tried two types of the preconditioner: (i) using diagonal elements of the kinetic energy
\begin{equation}
T_m^{(p,q)}(\mathbf{G})\equiv \left( \frac{1}{2}|\mathbf{k}+\mathbf{G}|^2 -\epsilon_m^{(q)} \right)^{-1},
\end{equation}
and (ii) an empirical form proposed by M. C. Payne {\it et al.}~\cite{Payne_precon} in the context of the CG method
\begin{equation}
T_m^{(p,q)}(\mathbf{G})\equiv \frac{27+18x+12x^2+8x^3}{27+18x+12x^2+8x^3+16x^4},
\end{equation}
where
\begin{equation}
x = \frac{1}{2}|\mathbf{k}+\mathbf{G}|^2 \left( \sum_{\mathbf{G'}} \frac{1}{2}|\mathbf{k}+\mathbf{G'}|^2 |v_m^{(p,q)}(\mathbf{G'})|^2 \right)^{-1}.
\end{equation}
We found that both schemes sufficiently improve the convergence of calculations. Here we used the latter one, but no large difference is observed for calculations in this study.
Then we perform the Gram-Schmidt orthonormalization for $\{ \tilde{v}_1^{(p+1,q)}, \tilde{v}_2^{(p+1,q)},\dots, \tilde{v}_n^{(p+1,q)} \}$ and obtain $\{ v_1^{(p+1,q)}, v_2^{(p+1,q)}, \dots, v_n^{(p+1,q)} \}$ where $v_m^{(p+1,q)}$ is orthogonalized with $v_i^{(j,q)}$ ($1\leq i \leq n$, $1\leq j \leq p$) and $v_i^{(p+1,q)}$ ($1\leq i \leq m-1$).
In other words, the Gram-Schmidt orthonormalization is performed for $n(p+1)$ vectors, $\{ \{v_1^{(1,q)}, v_2^{(1,q)}, \dots, v_n^{(1,q)}\}, \allowbreak \{ v_1^{(2,q)}, v_2^{(2,q)}, \dots, v_n^{(2,q)}\}, \allowbreak \dots, \{v_1^{(p,q)}, v_2^{(p,q)}, \dots, v_n^{(p,q)}\}, \allowbreak \{\tilde{v}_1^{(p+1,q)}, \tilde{v}_2^{(p+1,q)}, \dots, \tilde{v}_n^{(p+1,q)}\} \}$, and only $\tilde{v}$ are changed.
The subspace for diagonalization is now spanned with these $n(p+1)$ vectors.
Then we return to the beginning of the $p$ loop and continue to expand the subspace dimension up to $np_{\mathrm{max}}$.

\subsection{Outer loop: diagonalization and convergence check}

After the end of the inner $p$ loop, we evaluate the matrix elements of $F[v^{(1,\tilde{q})}]$ in the subspace spanned with $v_i^{(j,q)}$ ($1\leq i \leq n$, $1\leq j \leq p_{\mathrm{max}}$), and then diagonalize it.
The eigenvectors and eigenvalues obtained here are used as the initial guesses for the next $q$ loop: $\{ v_1^{(1,q+1)}, v_2^{(1,q+1)}, \dots, v_n^{(1,q+1)} \}$ and $\{ \epsilon_1^{(q+1)}, \epsilon_2^{(q+1)}, \dots,\epsilon_n^{(q+1)} \}$ where the only $n$ eigenvectors and eigenvalues are picked up from the lowest eigenvalue. 
Trial vectors $v$ included in $F[v^{(1,\tilde{q})}]$ are updated for every $N_{\mathrm{update}}$ loops of $q$, i.e., $\tilde{q}$ is defined as (the maximum multiple of $N_{\mathrm{update}}$) $+1$ not exceeding $q$.
At the same time as such update, we check convergence for $\tilde{E}^{(q)}\equiv \sum_i f(\epsilon_i^{(q)})\epsilon_i^{(q)}$ where $f(\epsilon)$ is the occupation number, instead of a bit expensive evaluation of the total energy in the TC method. $f(\epsilon)$ is always 0 or 1 in this study but can be a fractional number for metallic systems.
If we store the matrix elements of the one, two, and three-body terms in the Fock matrix separately, the total energy can be efficiently evaluated and future implementation along this line is possible, but the present convergence criteria does not arise any problem in this study.
For band structure calculation after an self-consistent determination of the occupied orbitals\cite{nonSCFnote}, we instead check convergence of $\sum_i \epsilon_i^{(q)}$.
If convergence is achieved, we once evaluate the total energy for output and calculation ends. Otherwise, the $q$ loop is continued.
In principle, other convergence criteria are also applicable, e.g., that with respect to the electron density.
All updates, evaluation of matrix elements, generation of new basis, the Gram-Schmidt orthonormalization, and diagonalization in our algorithm are performed simultaneously with respect to the $k$-points.

\subsection{Other issues}

We comment on the choice of the parameters.
In this study, we fixed $p_{\mathrm{max}}$ to be 2, i.e., the max value of the subspace dimension is $2n$. 
We also fixed $N_{\mathrm{update}}$ to be 3 for SCF calculations.
We verified that these choices work well both for accuracy and efficiency in this study.

For the BiTC method, we employ essentially the same algorithm, but the orthonormalization condition is replaced with the bi-orthonormalization conditions: $\langle  w_i^{(1,q)} | v_j^{(1,q)} \rangle = \delta_{ij}$ and $\langle v_i^{(1,q)} | v_i^{(1,q)} \rangle = 1$ ($1\leq i, j \leq n$) where $w$ are the left trial vectors~\cite{Davidson_nonsym}.
Note that the left and right eigenvectors are obtained simultaneously in the subspace diagonalization.

To improve the speed of convergence, we also employed the Pulay's scheme for density mixing~\cite{Pulay1,Pulay2}. The density mixing is performed when $v$ in $F[v]$ are updated.

\section{Results}\label{sec:4}
\subsection{Computational details}\label{sec:4A}
Norm-conserving Troullier-Martins pseudopotentials~\cite{TM} in the Kleinman-Bylander form~\cite{KB} constructed in LDA~\cite{PZ81} calculations were used in this study.
For Li, C, and Si atoms, we used two types of pseudopotentials for each: all-electron or [He]-core for Li and C, and [He]- or [Ne]-core for Si.
In this study, 1$s$ core states of Si are always included in the pseudopotential and not explicitly treated.
Even in `all-electron' calculations, we used small core radii (0.9 Bohr for H $s$ state, 0.4 and 0.8 Bohr for Li $s$ and $p$ states, and 0.2 Bohr for C $s$ and $p$ states, respectively) for non-local projection.
In this study, we call calculations using smaller-core pseudopotentials for each atomic species and those using larger-core ones `with-core' and `without-core' calculations, respectively. We call the occupied states in `without-core' calculations valence states hereafter.

Singularities of the electron-electron Coulomb repulsion and the Jastrow function in the $k$-space were handled with a method proposed by Gygi and Baldereschi~\cite{GygiBaldereschi} using an auxiliary function of the same form as Ref.~[\onlinecite{auxfunc}].
Static dielectric constants calculated in our previous study~\cite{TCjfo} using an RPA formula were adopted in determination of the Jastrow parameter $A$ (Eq.~(\ref{eq:JastrowA})).
Here we used the same value of the static dielectric constants independent of the treatment of core states for a fair comparison between with-core and without-core calculations.
Note that a value of $N$ (see Eqs.~(\ref{eq:Hamil}) and (\ref{eq:JastrowA})) is by definition different between with-core and without-core calculations.
Here, consistency of $N$ between Eq.~(\ref{eq:Hamil}) and Eq.~(\ref{eq:JastrowA}) is required to describe a screening effect through the three-body terms in the TC Hamiltonian in a way presented in our previous study~\cite{TCjfo}.

Experimental lattice parameters used in this study were also the same as those in Ref.~[\onlinecite{TCjfo}].
LDA calculations were performed with \textsc{tapp} code~\cite{tapp1,tapp2} to make an initial guess for TC orbitals. Subsequent TC calculations were performed with \textsc{tc}{\small ++} code~\cite{Sakuma,TCaccel}. We also used LDA orbitals as basis functions for TC orbitals in Sec.~\ref{sec:4B} to compare its performance with that for calculation using a plane-wave basis set. These calculations are called LDA-basis and plane-wave-basis calculations hereafter.

\subsection{Efficiency of plane-wave-basis calculation}\label{sec:4B}

\begin{figure}
 \begin{center}
  \includegraphics[width=8 cm]{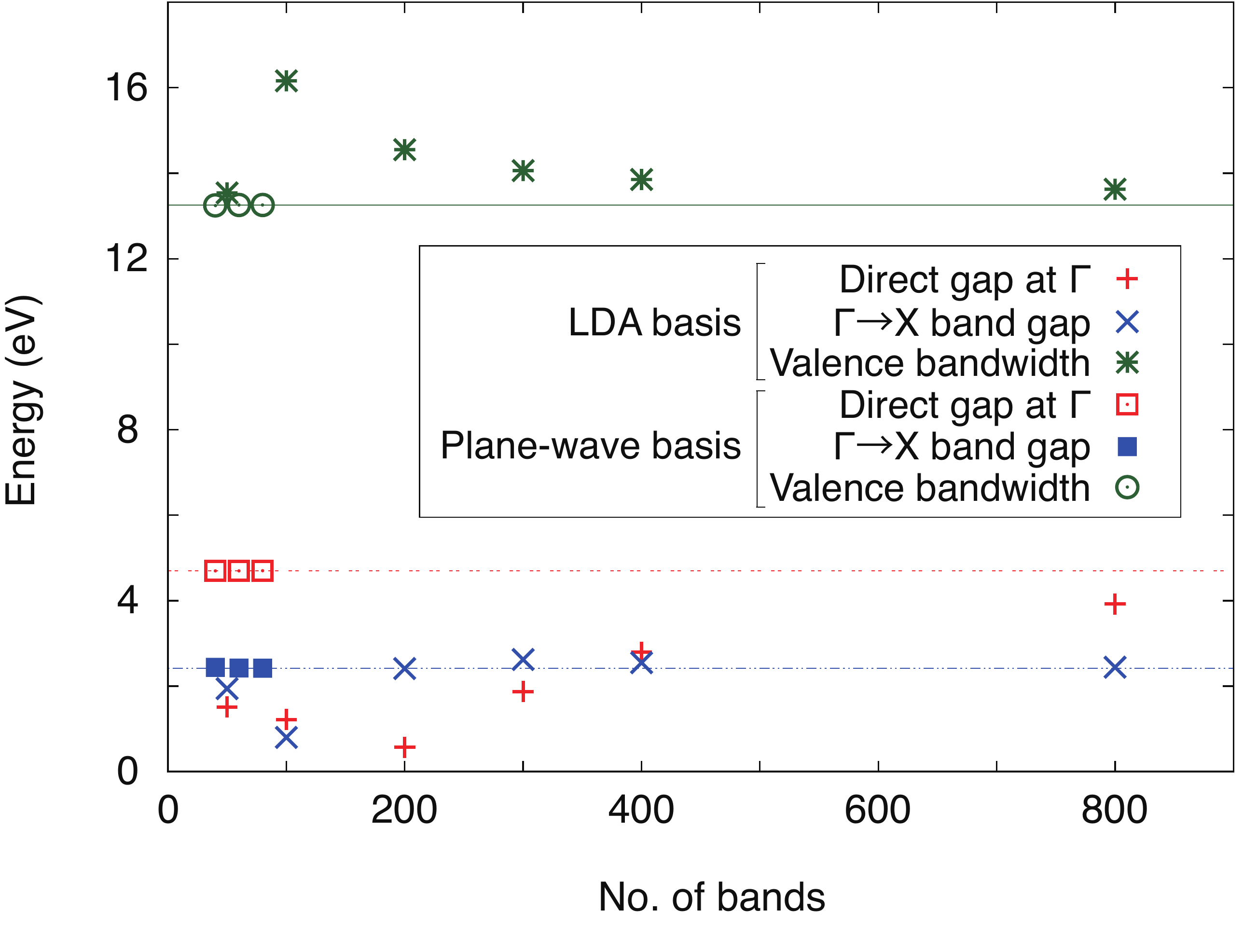}
  \caption{Convergence of the band structure for bulk silicon with core states in terms of the number of bands is presented for LDA-basis and plane-wave-basis calculations. Lines show the values corresponding to the rightmost data points for plane-wave-basis calculations as guides to the eyes.}
  \label{fig:bandconv}
 \end{center}
\end{figure}

\begin{figure}
 \begin{center}
  \includegraphics[width=8 cm]{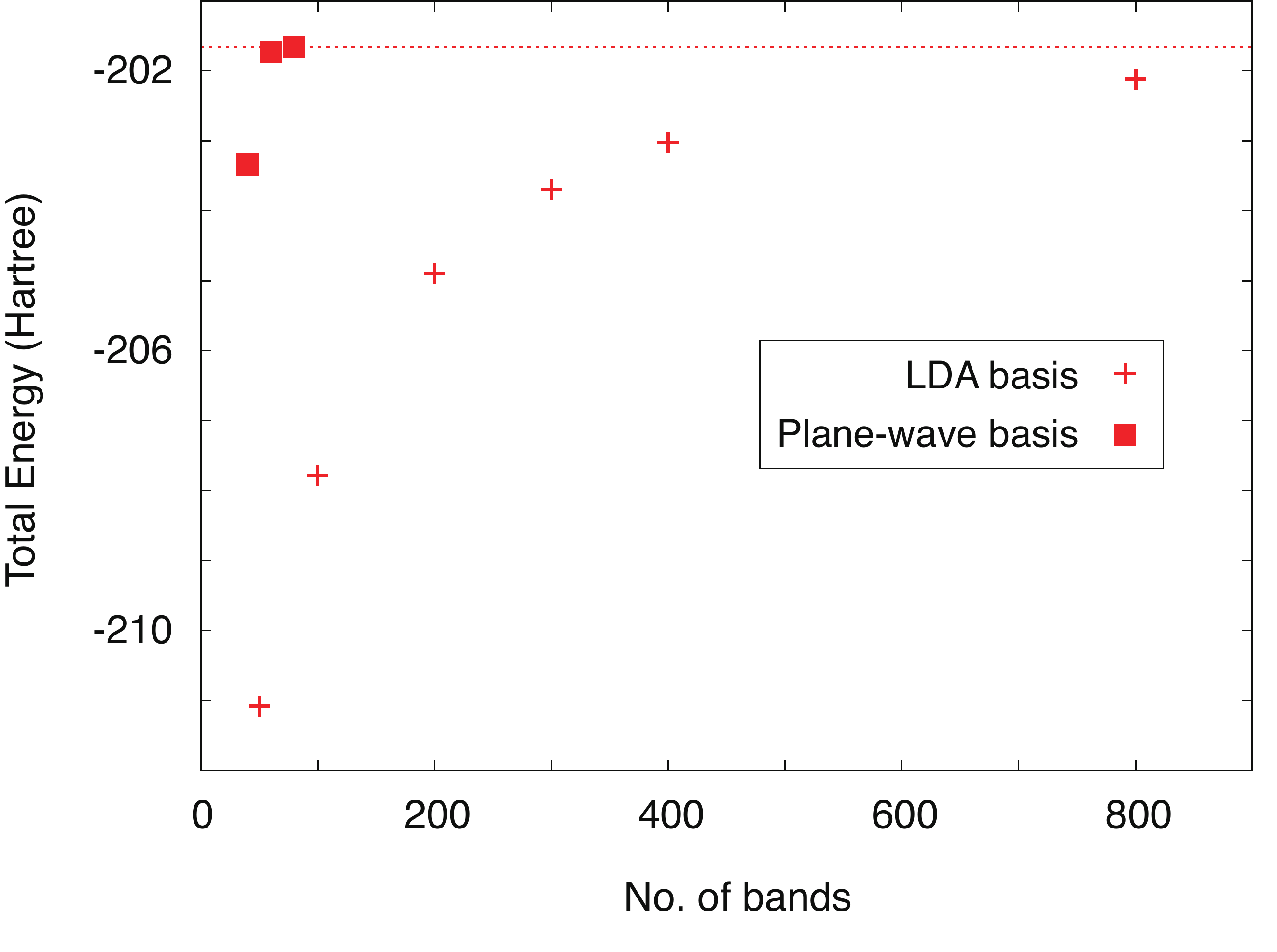}
  \caption{Convergence of the total energy for bulk silicon with core states in terms of the number of bands is presented for LDA-basis and plane-wave-basis calculations. A line shows the value corresponding to the rightmost data point for plane-wave-basis calculations as a guide to the eyes.}
  \label{fig:Econv}
 \end{center}
\end{figure}

To see how efficiently our new algorithm works, we performed band structure calculations for bulk silicon with core states. This is because a regular setup, bulk silicon without core states,  requires only a small number of subspace dimension both for LDA-basis and plane-wave-basis calculations and is not appropriate as a test case here.
In Figure~\ref{fig:bandconv}, we present the convergence behavior of the band structure for bulk silicon with core states in terms of the number of bands $n$, where the subspace dimension for diagonalization is $n$ and $np_{\mathrm{max}}=2n$ for LDA-basis and plane-wave-basis calculations, respectively (see Section~\ref{sec:3}).
Band gaps between the highest valence and lowest conduction bands at the $\Gamma$ point, that between the highest valence band at the $\Gamma$ point and the lowest conduction band at the $X$ point, and the valence bandwidth at the $\Gamma$ point are shown here.
A cut-off energy of plane waves was 256 Ry, and a $2\times 2\times 2$ $k$-point mesh was used throughout this subsection.
Convergence of the total energy is also shown in Figure~\ref{fig:Econv}.

In these figures, we can see that LDA-basis calculations require an enormously large number of the subspace dimension. Even when we used 800 LDA bands, the direct gap at the $\Gamma$ point still exhibits an error of about 0.8 eV compared with the well-converged value obtained in plane-wave-basis calculation.
Total energy also shows much slower convergence in LDA-basis calculation than in plane-wave-basis calculation.
Because one needs to take such a huge number of LDA bands even for bulk silicon when one explicitly takes the core states into account, LDA-basis calculation for more complex materials such as strongly correlated systems seems to be intractable both in computation time and memory requirement.
On the other hand, plane-wave-basis calculations require a moderate number of bands to achieve sufficient convergence especially for the band structure calculations.

\begin{figure}
 \begin{center}
  \includegraphics[width=8 cm]{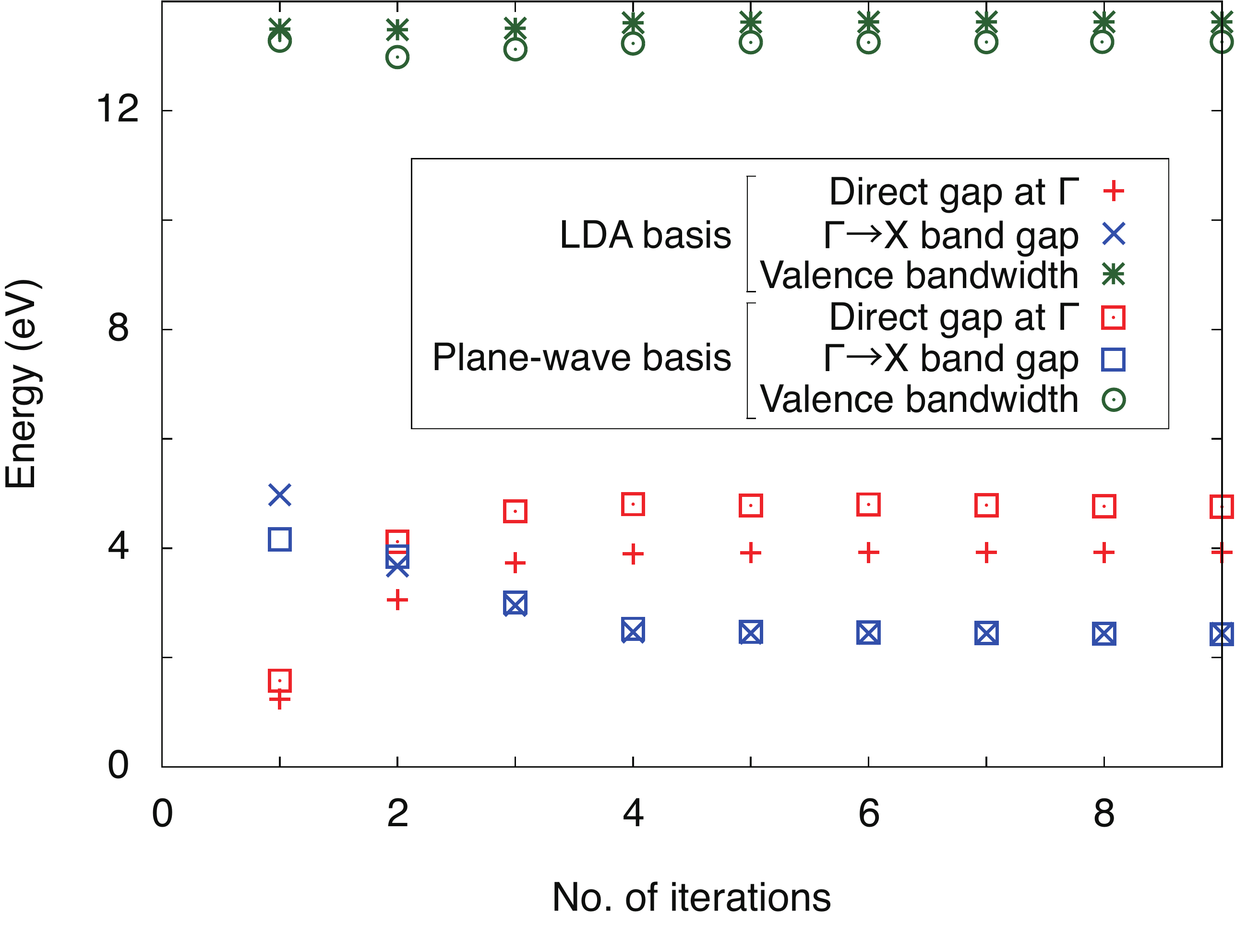}
  \caption{Convergence of the band structure for bulk silicon in terms of the number of iterations is presented for LDA-basis ($n=800$) and plane-wave-basis ($n=60$) calculations.}
  \label{fig:banditer}
 \end{center}
\end{figure}

Figure~\ref{fig:banditer} presents the convergence behavior of the band structure with respect to the number of iterations for LDA-basis and plane-wave-basis calculations.
LDA-basis and plane-wave-basis calculations shown here were performed with the number of bands $n$ being 800 and 60, respectively, where the convergence errors in calculations with two basis sets are comparable using these values of $n$ as we have seen in Figs.~\ref{fig:bandconv} and \ref{fig:Econv}.
The number of iteration here denotes that for the outer loop described in Section~\ref{sec:3} for plane-wave-basis calculations.
Since we did not calculate the total energy at each iteration in plane-wave-basis calculations as noted in Section~\ref{sec:3}, we only show data for the band structure.
We verified that about 5 iterations are sufficient to obtain convergence within an error of 0.1 eV for both basis sets in this case.
As for the computation time, one iteration takes about 60 and 22 hours without parallelization on our workstation for LDA-basis and plane-wave-basis calculations, respectively.
Note that LDA-basis calculation with the number of bands $n=800$ still does not achieve sufficient convergence as mentioned in the previous paragraph, which means that, to achieve the same level of accuracy, LDA-basis calculation needs much longer computation time and larger memory requirement than plane-wave-basis calculation.

\subsection{Application to band structure calculations with core states}

As a test of our new algorithm, we investigated the band structures with core states using a plane-wave basis set.
Because accurate calculation of the total energy with an explicit treatment of core electrons requires sizable computational effort, we concentrate on the band structures in this study.

\begin{figure}
 \begin{center}
  \includegraphics[width=8.5 cm]{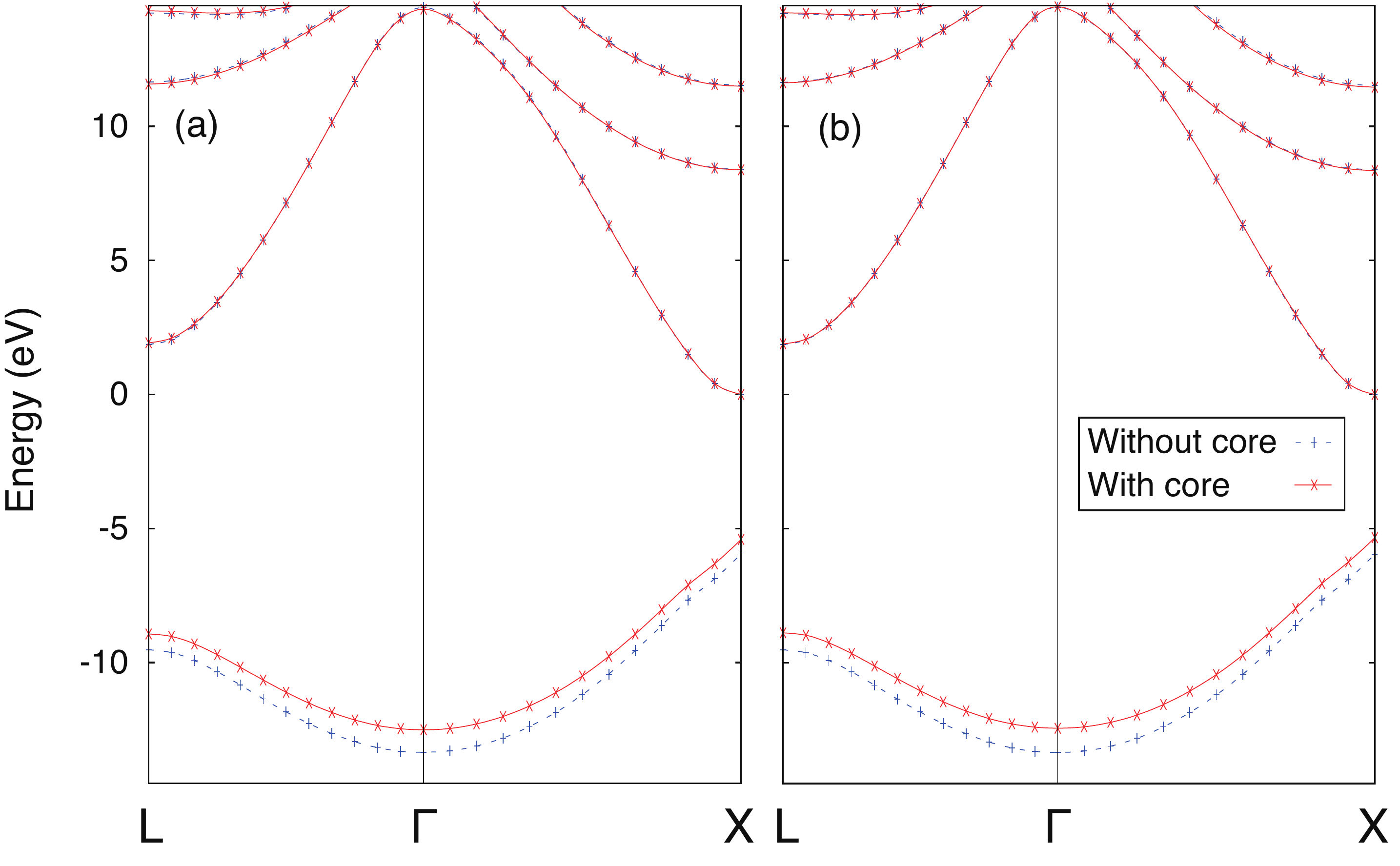}
  \caption{(a) TC and (b) BiTC band structures of LiH with (red solid lines) and without (blue broken lines) an explicit treatment of the Li-1$s$ core states. Conduction band bottoms for both band dispersions are set to zero in the energy scale.}
  \label{fig:LiH}
 \end{center}
\end{figure}

In Figure~\ref{fig:LiH}, we present the band structures of LiH calculated with the TC and BiTC methods using a $6\times 6\times 6$ $k$-mesh and $n=20$.
We used 49 and 169 Ry for the cutoff energy of plane waves in without-core and with-core calculations, respectively.
There is almost no difference between the TC and BiTC band structures for this material.
We can see that the position of the valence band is affected by inclusion of core electrons, which improves the agreement of the calculated band gap with experimental one as presented in Table~\ref{table:band}~\cite{noteLiH}.
Band structures in the upper energy region are almost unchanged between with-core and without-core calculations.

\begin{figure}
 \begin{center}
  \includegraphics[width=8.5 cm]{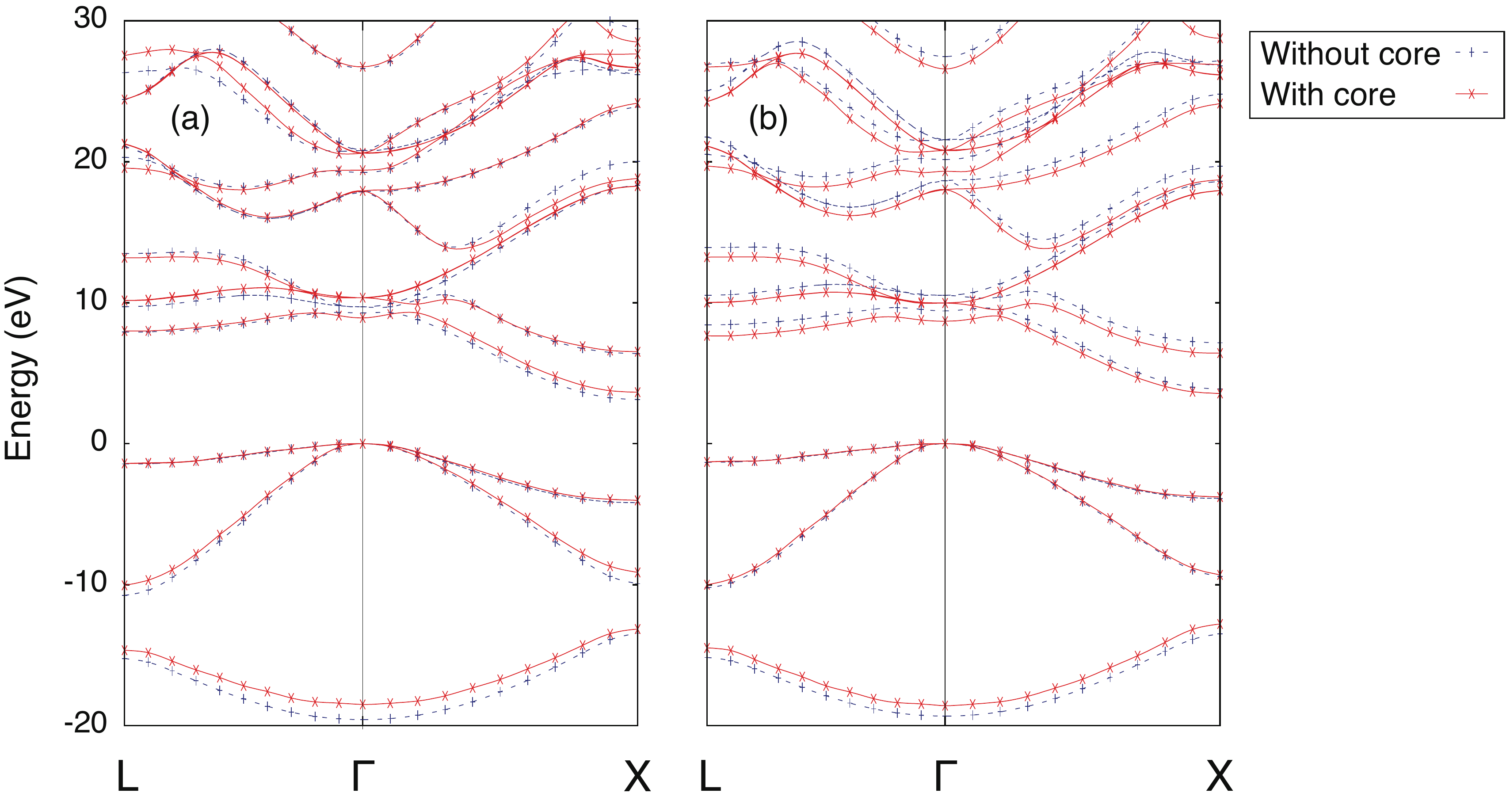}
  \caption{(a) TC and (b) BiTC band structures of $\beta$-SiC with (red solid lines) and without (blue broken lines) an explicit treatment of the C-1$s$ and Si-2$s$2$p$ core states. Valence-band tops for both band dispersions are set to zero in the energy scale.}
  \label{fig:SiC}
 \end{center}
\end{figure}

We also calculated the band structures of $\beta$-SiC as shown in Figure~\ref{fig:SiC} using a $4\times 4\times 4$ $k$-mesh.
Cutoff energies of 121 and 900 Ry and $n=30$ and $40$ were used for without-core and with-core calculations, respectively.
We again see that an explicit inclusion of core electrons makes the position of the deepest valence band a bit shallow.
A similar trend is found also for Si, where the overestimated valence bandwidth is improved by an inclusion of core states as presented in Table~\ref{table:band}.
This feature can be relevant to the observed overestimation of the valence bandwidth for bulk silicon calculated with the state-of-the-art diffusion Monte Carlo (DMC) method~\cite{SiDMC}.
In the upper energy region, band structures of $\beta$-SiC are again almost unchanged but exhibit some differences between with-core and without-core calculations.
The difference in the BiTC band structures can be interpreted as mere shifts for each angular momentum.
The TC band structures, on the other hand, show changes that are not constant shifts.
Note that these slight changes are natural because our Jastrow factor is different between with-core and without-core calculations as mentioned before.
Furthermore, the LDA core states used to construct our LDA pseudopotentials should be different from those in the TC and BiTC methods, which naturally can, in principle, produce complicated changes in the band structures.
In addition, the deep core states explicitly included in the many-body wave function are expected to dominate the wave-function optimization, which can affect the obtained eigenvalues in the upper energy region.
It is noteworthy that, nevertheless, the BiTC band structures are always improved by an inclusion of deep core states as shown in Table~\ref{table:band}, for materials we calculated here. 
The position of the deepest valence bands is always improved both for the TC and BiTC methods.
Further investigation for different materials is an important future issue.

It is also an important future issue to construct (Bi)TC pseudopotentials that are free from the aforementioned pseudopotential errors evaluated in our analysis. Such development is a nontrivial task because of the difference of the Jastrow factor among with/without-core calculations for solids and atomic calculations required in constructing the pseudopotentials, but the similarity between the band structures obtained in with-core and without-core calculations is a hopeful observation for this purpose.

\begin{table*}
\begin{center}
\begin{tabular}{c c c c c c c c c}
\hline \hline
 & & \multicolumn{2}{c}{LDA} & \multicolumn{2}{c}{TC} & \multicolumn{2}{c}{BiTC}  & Expt.\\
& with core & N & Y & N & Y & N & Y  & -\\ 
\hline
LiH & Band gap & 2.6 & 2.6 & 6.6 & 6.0 & 6.6 & 5.9 & 5.0$^a$\\
 & Valence bandwidth & 5.5 & 5.3 & 6.8 & 6.5 & 6.8 & 6.5 & 6.3$\pm$1.1$^b$\\
 $\beta$-SiC & Indirect band gap & 1.3 & 1.3 & 3.1 & 3.6 & 3.9 & 3.6 & 2.4$^c$ \\
 & Direct band gap & 6.3 & 6.3 & 9.3 & 8.9 & 9.4 & 8.7 & 6.0$^d$ \\
 & Valence bandwidth & 15.3 & 15.3 & 19.6 & 18.5 & 19.3 & 18.6 & - \\
Si & Indirect band gap & 0.5 & 0.5 & 2.0 & 2.2 & 2.2 & 1.8 & 1.17 $^e$\\
 & Direct band gap & 2.6 & 2.5 & 4.6 & 4.5 & 4.6 & 3.6 & 3.40, 3.05$^f$ \\
 & Valence bandwidth & 11.9 & 11.9 & 15.1 & 13.5 & 15.1 &14.2 & 12.5$\pm$0.6$^f$\\
\hline \hline
\end{tabular}
\end{center}
\caption{\label{table:band} Band energies calculated with LDA, TC, and BiTC methods for LiH, $\beta$-SiC, and Si. All in eV.
$^a$ Reference \onlinecite{LiHbandExpt}.
$^b$ Reference \onlinecite{LiHbwdthExpt}.
$^c$ Reference \onlinecite{SiCbandExpt}.
$^d$ Reference \onlinecite{SiCdbandExpt}.
$^e$ Reference \onlinecite{SibandExpt}.
$^f$ From the compilation given in Reference \onlinecite{SidbandExpt}.}
\end{table*}

\section{Summary}\label{sec:5}

In this study, we develop an iterative diagonalization scheme for solving an SCF equation of the TC method using a plane-wave basis set.
We make use of the block-Davidson algorithm and verify that our new scheme effectively reduces computational requirement both in memory and time.
Also an influence of the core states in the TC calculations is investigated.
We find that an explicit treatment of core states improves a position of deep valence states whereas the band structures in upper energy region do not exhibit large changes, which means that our choice of the Jastrow factor can provide consistent electronic structures for a wide energy range whether the core electrons are explicitly treated or not.
Our study opens the way to further application of the TC method to the strongly correlated systems.

\section*{Acknowledgments} 

This study was supported by a Grant-in-Aid for young scientists (B) (Number 15K17724) from the Japan Society for the Promotion of Science, MEXT Element Strategy Initiative to Form Core Research Center, and Computational Materials Science Initiative, Japan.


\begin{thebibliography}{999}
\bibitem{nonSCFnote} In the TC method, band structure calculation after the regular SCF calculation is performed self-consistently. This is because the singularities of the interaction in the $k$-space were handled in the manner that requires the one-body wave functions on the $k$-mesh used in the band-structure plot.
\bibitem{noteLiH} For LiH, note that our LDA pseudopotentials do not include the non-linear core correction~\cite{NCC} since its meaning is ambiguous in the context of the TC method. This can be an origin of a difference for the valence bandwidths obtained by with-core and without-core LDA calculations presented in Table~\ref{table:band}. For other materials, the calculated LDA band structures are almost unchanged between with-core and without-core calculations.
\end{thebibliography}
\end{document}